\documentclass[fleqn,10pt]{wlpeerj} 

\newcommand{\beginsupplement}{%
        \setcounter{table}{0}
        \renewcommand{\thetable}{S\arabic{table}}%
        \setcounter{figure}{0}
        \renewcommand{\thefigure}{S\arabic{figure}}%
     }
     
\begin{document}

\title{Paradoxes in Leaky Microbial Trade}

\author[1]{Yoav Kallus}
\author[1,2]{John H. Miller}
\author[1,*]{Eric Libby}
\affil[1]{Santa Fe Institute, 1399 Hyde Park Road, Santa Fe, NM, USA, 87501}
\affil[2]{Carnegie Mellon University, Pittsburgh, PA, USA, 15213}
\corrauthor[*]{Eric Libby}{elibby@santafe.edu}

\begin{abstract}
Microbes produce metabolic resources that are important for cell growth yet leak across membranes into the extracellular environment. Other microbes in the same environment can use these resources and adjust their own metabolic production accordingly---causing other resources to leak into the environment. The combined effect of these processes is an economy in which organismal growth and metabolic production are coupled to others in the community. We propose a model for the co-evolving dynamics of metabolite concentrations, production regulation, and population frequencies for the case of two cell types, each requiring and capable of producing two metabolites. In this model, beneficial trade relations emerge without any coordination, via individual-level production decisions that maximize each cell's growth rate given its perceived environment. As we vary production parameters of the model, we encounter three paradoxical behaviors, where a change that should intuitively benefit some cell type, actually harms it. (1) If a cell type is more efficient than its counterpart at producing a metabolite and becomes even more efficient, its frequency in the population can decrease. (2) If a cell type is less efficient than its counterpart at producing a metabolite but becomes less inefficient, the growth rate of the population can decrease. (3) Finally, if a cell type controls its counterpart's production decisions so as to maximize its own growth rate, the ultimate growth rate it achieves can be lower than if the two cell types each maximized their own growth. These three paradoxes highlight the complex and counter-intuitive dynamics that emerge in simple microbial economies. 
\end{abstract}

\flushbottom
\maketitle
\thispagestyle{empty}

\section*{Introduction}

Microbes live in complex communities where goods such as metabolites are produced and exchanged \cite{Nadell:2016bl,Sachs:2012hn,Kouzuma:2015fj,Johns:2016kc}. As goods flow in and out of cells, a type of economy emerges \cite{Werner:2014dy,Tasoff:2015id}. In this economy, each organism faces decisions concerning which goods to produce and in what quantities \cite{Hammerstein:2016bn}. These production decisions ultimately determine the relative abundance of each organism since more successful individuals will grow faster and increase in frequency. As populations change, the economic conditions can change and put pressure on organisms to adjust their production \cite{Tasoff:2015id}. In this paper, we investigate this interplay between population-level dynamics and individual-level production decisions and uncover paradoxical system-level behaviors.

Microbes exchange goods directly or indirectly \cite{Morris:2015dq}. Direct mechanisms, such as intercellular nanotubes \cite{Pande:2015hb} or cell-cell recognition systems \cite{Kouzuma:2015fj}, allow microbes to target goods towards specific partners, thereby facilitating successful trading relationships. In contrast, indirect exchange typically relies on the diffusion of molecules through the extracellular environment \cite{Morris:2015dq,Estrela:2016kn}. Some goods are produced and secreted because their primary function occurs extracellularly. One classic example is a siderophore that binds extracellular iron and allows it to be imported into the cell \cite{Cordero:2012bi}. Other goods diffuse out of cells through inherently permeable cell membranes \cite{Morris:2015dq}. Metabolic byproducts and electron carriers are example of these kinds of leaked goods \cite{Morris:2015dq,Schink:2002uy}. Once such goods are in the environment they can be used to inform individual production strategies. Here, we focus exclusively on indirect exchange of goods via diffusion.

Even if we consider only indirect exchange of diffusible goods, there is a great diversity of types of exchange depending on the environmental and ecological context, the number of organisms and goods, as well as the costs and benefits of the goods \cite{Estrela:2016kn,Morris:2013ja}. We narrow our scope by considering only interactions between two organisms involving two goods. This excludes well-studied systems of trade such as the mutualism between mycorrhizal fungi and plants in which many different organisms may be trading simultaneously  \cite{Wyatt:2014jv,Kummel:2006bl}. Furthermore, we only consider goods that are costly to make and beneficial to at least one organism. Thus, we do not consider punitive goods such as toxins or antibiotics. We find it useful to classify goods in terms of which organisms produce them and which organisms benefit from their consumption. Using this approach,  Table \ref{table-models} shows three canonical types of exchange. For each good, we denote which organism produces it (\textit{p}) and which organism consumes it (\textit{c}). The three types of exchange do not represent an exhaustive classification, but rather provide a way of comparing exchange interactions that have received significant attention in previous studies.

\begin{table}[t]
\begin{center}
\begin{tabular}{r|cc|cc|cc}
& \multicolumn{2}{c|}{mutualism} & \multicolumn{2}{c|}{exploitation} & \multicolumn{2}{c}{self-sufficiency} \\
& organism 1 & organism 2 & organism 1 & organism 2 & organism 1 & organism 2  \\
\hline 
good 1 & p & c & p,c & c & p,c & p,c \\
good 2 & c & p & --- & p,c & p,c & p,c \\
\end{tabular}
\caption{Classification of microbial exchange between two organisms involving two goods. A \textit{p} indicates that a good is produced and a \textit{c} indicates that it is consumed. We assume that if an organism consumes a good, it benefits in some way.}
\label{table-models}
\end{center}
\end{table}

The first category is {\it mutualism}, where each organism produces a good that the other one consumes. This type of relationship can represent syntrophy \cite{McInerney:2008hp,Morris:2013ja}, cross-feeding \cite{Bull:2009dr}, auxotrophy \cite{Johns:2016kc}, or a two-way byproduct mutualism \cite{Eberhard:1975hi,Sachs:2004fe}. Since each organism does not consume the good that they produce, the goods are byproducts of other processes. This means that the optimal amount of the byproduct to produce depends on the costs and benefits of the other, more primary processes as well as how much benefit is derived from the good produced by the other organism \cite{Sachs:2004fe,Bull:2009dr,Doebeli:1998uo}. In instances where each good produced is growth-limiting to the other organism, there is a positive feedback loop so each organism does best by producing as much of their good as possible so long as it does not interfere with other cell functions. One common result of these syntrophic exchanges is a synergy between organisms, where the combined community has enhanced growth relative to any isolated individual \cite{Tasoff:2015id}. 

In the second category, {\it exploitation}, one organism produces a good that both organisms value, while the other organism produces only goods of value to itself. This arrangement captures parasitic behaviors as well as forms of cheating and competition \cite{Ghoul:2016cz,Bull:2009dr}. Indeed, this arrangement describes the public goods dilemma that has been well-studied in social evolution \cite{Cordero:2012bi}. Although one organism is exploiting the other, there is no real production decision for the producer since it needs the good and is the only one that produces it. This situation is at the heart of the Black Queen hypothesis, where adaptive gene loss leaves one organism burdened with producing a costly metabolite that is exploited by the community \cite{Morris:2012bd}.

The final category, {\it self-sufficiency}, represents the most flexible and possibly primitive arrangement. Here, each organism is capable of producing all of the goods it needs for survival and both organisms value these goods. Possible goods that fit this scenario include amino acids or molecules essential to central metabolism or maintenance. Interestingly, this category is a precursor to the other categories, as loss of function mutations can result in either mutualism or exploitation scenarios. Thus, we focus exclusively on this arrangement in order to understand how its dynamics might prime populations to evolve into one of the other categories.

The self-sufficiency case has been studied implicitly in models of metabolic trade. In these models, metabolic networks that are capable of growing on a variety of resources, are joined together to understand how the combined metabolism might function \cite{Biggs:2015gi,Stolyar:2007jh,Harcombe:2014it}. The production decisions are solved using some objective function and flux balance analysis. By joining metabolisms, it has been shown that extant organisms can grow on a wide variety of resources \cite{Klitgord:2010cs}. One feature lacking in these models is the dynamic interplay between population composition and production---especially when organisms have different production capabilities and there is a tension between maximizing individual and population growth rates. 

Here, we address the issue of population composition and growth with a general microbial trade model that couples population dynamics to organism production strategies. We assume that each organism alters its production in order to maximize its own growth rate. Since microbes can shift production of costly goods depending on environmental concentrations, each organism's production of leaky goods affects the production strategies of other organisms. Using this approach, we uncover three unusual system-level behaviors that apply to relevant trading scenarios between microorganisms. Furthermore, these behaviors suggest evolutionary trajectories that lead populations to more structured forms of arrangement such as mutualisms or exploitations.

\section*{Methods}

We consider a microbial population model in which organisms trade through the production and diffusion of metabolites. For simplicity, we assume there are two types of organisms ($1$ and $2$) that require the same two metabolites ($A$ and $B$) in order to grow and reproduce. We denote the amount of $A$ and $B$ metabolites in cells of type $i = 1,2$ as $A_i$ and $B_i$ and the number of cells of type $i$ by $N_i$. We assume that the population growth rate is proportional to a rate $g_i(A_i,B_i)$ determined by the internal concentration of the metabolites in each organism. Although there may be many possible growth functions $g_i$, we choose the general functional form
\begin{linenomath*}
\begin{equation}\label{eq:G}
    g_i(A_i,B_i) = k_i A_i B_i\text.
\end{equation}
\end{linenomath*}
This represents a mass action law for an elementary reaction, wherein $A$ and $B$ react to form a product used directly for growth. 
We consider the simple case of the growth functions $g_i$ in which both organisms have the same growth function and $k_i=1$. This assumption implies that the organisms have similar metabolic needs. As a consequence of the growth process, metabolites $A$ and $B$ are consumed at rates $s_{A,i} g_i(A_i,B_i)$ and $s_{B,i} g_i(A_i,B_i)$, respectively. The stoichiometry coefficients, $s_{A,i}$ and $s_{B,i}$, depend on the growth reaction and here we investigate the simple case where $s_{A,i}=s_{B,i}=1$.

Since we choose to analyze the self-sufficiency case in Table \ref{table-models}, each organism can produce both $A$ and $B$ metabolites. Production, however, comes with costs either as a result of energy expenditure or forfeited opportunities to produce other goods or engage in other processes. We assume that the production rates of metabolites are subject to a budget constraint whereby the organism has a finite amount of resources (precursors, enzymes, ribosomes, etc.)\ that can be devoted to the production of metabolites. We encapsulate all the relevant constraints in the production constraint function, $P_i(p_{A,i},p_{B,i})$, subject to a constraint $P_i(p_{A,i},p_{B,i})\le P_\text{max}$ where $p_{X,i}$ is the rate of production of metabolite $X$ by cells of type $i$. For example, 
\begin{linenomath*}
\begin{equation}\label{eq:P}
    P_i(p_{A,i},p_{B,i}) = c_{A,i} p_{A,i} + c_{B,i} p_{B,i} \le 1
\end{equation}
\end{linenomath*}
represents a situation where metabolites $A$ and $B$ can be produced at fixed costs ($c_{A,i}$ and $c_{B,i}$, given in units of the total budget) independent of the total rate of production. Thus, there are no returns to scale.

Besides consumption and production, metabolites can be gained or lost through passive diffusion depending on the concentration gradient across the cell membrane. We assume that there is a rate of diffusion of the metabolite molecules out of any cell and into a random other cell. The total flux of molecules leaving cells of any type will be proportional to a diffusion coefficient $D$, the intracellular concentrations of the molecules, and the number of cells of this type. Since diffusion is unbiased in our model, molecules will enter cells of type $1$ or $2$ according to their proportions in the population. We define the relative frequency of cells of type $i$ by $n_i = N_i/(N_1+N_2)$. As a result, the net flux of $A$ molecules entering a single cell of type $1$ is $D n_2 (A_2 - A_1)$ and similarly $D n_1 (A_1 - A_2)$ for cells of type $2$. The diffusion coefficient $D$ determines the relative rate at which molecules flow down a gradient as opposed to getting consumed by the growth reaction. Therefore, the smaller $D$ is, the more benefit a microbe derives from producing a metabolite directly as opposed to relying on a trading partner. We use $D=3$ in the numerical cases investigated in the main text of the paper, but show the effects of varying $D$ in the supplementary material. 

In addition to cross-cell diffusion, we assume that there is a rate $\mu$ at which metabolites are lost and not regained by any cell, either due to diffusion away from the shared environment or by some process of degradation. We set this loss rate to be $\mu=0.05$ throughout the paper. Although the precise value of $\mu$ does not change the key results of the paper, we analyze the effects of varying $\mu$ in the supplementary material.

These dynamical processes result in a set of differential equations that describes the intracellular concentrations of $A$ and $B$ metabolites in the two types of cells:
\begin{linenomath*}
\begin{equation}\label{eq:dXidt}
    \begin{aligned}
	\frac{dA_1}{dt} &= p_{A,1} + D n_2 (A_2 - A_1) -\mu A_1 - s_{A,1} g_1(A_1,B_1) \text,\\
	\frac{dB_1}{dt} &= p_{B,1} + D n_2 (B_2 - B_1) -\mu B_1 - s_{B,1} g_1(A_1,B_1) \text,\\
	\frac{dA_2}{dt} &= p_{A,2} + D n_1 (A_1 - A_2) -\mu A_2 - s_{A,2} g_2(A_2,B_2) \text,\\
	\frac{dB_2}{dt} &= p_{B,2} + D n_1 (B_1 - B_2) -\mu B_2 - s_{B,2} g_2(A_2,B_2) \text.
    \end{aligned}
\end{equation}
\end{linenomath*}
This dynamical system is similar to the one Taillefumier \textit{et al.}\ \cite{Taillefumier2016} use to study coordination among bacterial populations when exposed to a diverse resource supply. In our situation, we have no externally supplied metabolites and do not allow cells to interconvert metabolites: all metabolites are immutable and produced by the cells themselves. We also simplify the system by modeling diffusion between cells rather than explicitly modeling the extracellular environment. As a result, our system has three types of co-evolving dynamic variables: the intracellular metabolite concentrations ($A_i$ and $B_i$), the production terms ($p_{A,i}$ and $p_{B,i}$), and the relative population frequencies of the cell types ($n_1$ and $n_2$).

We assume that the dynamics with which the population sizes, $N_1(t)$ and $N_2(t)$, evolve is much slower than the rates of metabolite production and diffusion. At shorter time scales, the metabolite concentrations reach a steady state, where the time derivatives on the left hand sides of Eq.\ \eqref{eq:dXidt} equal zero. For particular values of the production rates, the steady-state values of the growth functions, denoted $g^*_1$ and $g^*_2$, can be determined by solving the resulting algebraic equations. If $g^*_1 > g^*_2$, then $n_1$, the relative frequency of cells of type 1, grows, thereby altering Eq.\ \eqref{eq:dXidt}. Since an increased $n_1$ affects the values of the steady-state growth rates, we then re-solve for the steady state with the increased $n_1$. This iterative process continues until a stable population equilibrium is reached. In order for the system to be in a stable equilibrium with both cell types at nonzero frequency, the steady-state growth rates must be equal, i.e. $g^*_1=g^*_2$. Alternatively, there could be an equilibrium where one cell type has a higher growth rate, while the other cell type is driven to a relative frequency approaching zero.

We have not yet discussed how the production rates evolve subject to the budget constraint. One possibility is to assume that the organisms can regulate these production rates on a fairly short time scale and cells of each type adjust their own production rates so as to maximize their own growth, subject to the perceived external conditions. This assumption leads to a situation where each cell type's choice of production rates is the best response to the external conditions, which are the result of the choice of the other cell type, implying a Nash equilibrium. While the assumption that the production can be regulated on a shorter time scale than the population dynamic time scale is convenient, it is not necessary: even if regulation only occurs through mutations, the system will be driven to a Nash equilibrium by the fact that a population not using the best-response production rates is susceptible to invasion by a mutation that does (see supplementary material). Computer code for our analyses is provided as supplementary files.

\section*{Results}
\subsection*{Extinction and coexistence}
Before we explore the behavior of interacting microbial populations, we first consider the growth of a population of cell types in the absence of trade. We prevent different cell types from exchanging metabolites by setting $D=0$ in Eq.\ \eqref{eq:dXidt}. We assume that every cell regulates its production rates $p_{A,i}$ and $p_{B,i}$ so as to maximize its own growth. Because our production constraint function Eq.\ \eqref{eq:P} does not feature returns to scale, there is no benefit to a division of labor among members of the same cell type. As a consequence, every cell of the same type shares an identical strategy in terms of how much of each metabolite is produced. We compute the growth rate for a cell type as a function of the energetic costs of making $A$ and $B$ metabolites, or equivalently,
the inverse of the costs. We call the inverse of a production cost the \textit{efficiency}, i.e.\ $a_{X,i}=1/c_{X,i}$, and it corresponds to the maximum amount of the good a cell can produce.

\begin{figure}
    \begin{center}
    \includegraphics[width=0.99\linewidth]{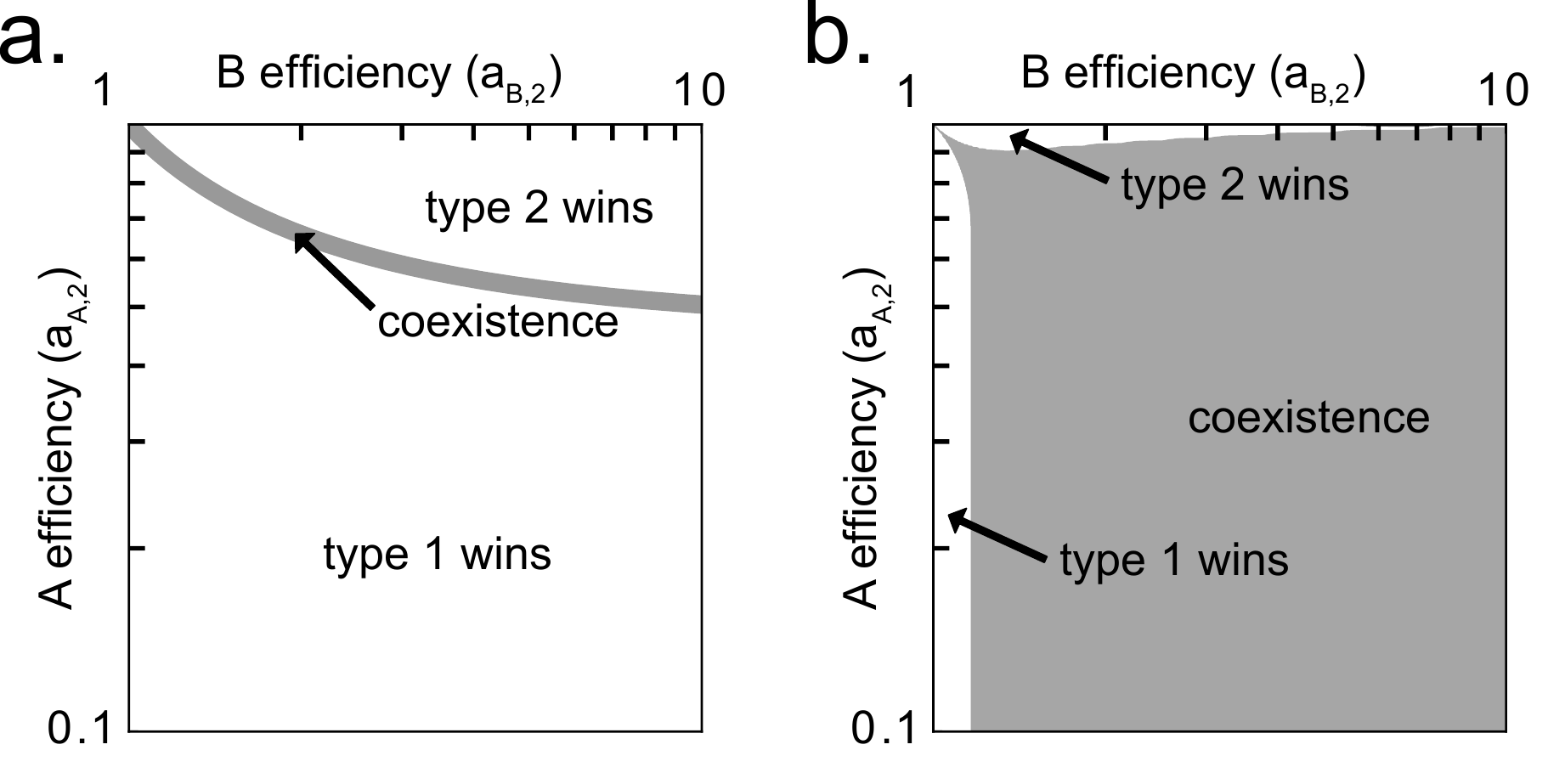} 
    \end{center}
    \caption{{\bf Coexistence with and without diffusion.} \textbf{a)} The growth rate of a population consisting of a single cell type is a function of the maximum amount of each essential metabolite that it can produce, i.e. $a_A$ for $A$ and $a_B$ for $B$. The gray line is the locus where the growth rate of a cell (type 2) equals that of a reference cell (type 1) that has equal costs for producing either metabolite, with $a_{A,1}=a_{B,1}=1$. In the absence of diffusion, coexistence with the reference cell type is only possible where the growth rates are equal. Above the gray line the type 2 cell grows faster and drives the type 1 cells to 0 relative frequency, i.e. the type 2 cells ``win.'' Below the line the situation is reversed. \textbf{b)} With diffusion, e.g., when $D=3$, each cell type population is affected by the other's production. The coexistence region is significantly larger and fills much of the quadrant considered, corresponding to where each cell type is more efficient than the other at producing one of the metabolites.
\label{fig:alone} }
\end{figure}

In Figure \ref{fig:alone}a, the gray line shows production efficiencies that yield the same growth rate as a reference cell type, say cell type 1, that is equally efficient at producing either metabolite, with $a_{A,1}=a_{B,1}=1$. When we add cell type 2 with production efficiencies given by $a_{A,2}$ and $a_{B,2}$ to a population of the reference cell type, one of the two cell types will grow faster and tend to 100\% of the population. Type 2 cells with efficiencies above the gray line in Figure \ref{fig:alone}a grow faster than the reference cell type and will ultimately drive it extinct; below the gray line, the reverse is true. Thus, in the absence of diffusion, coexistence is only possible on the gray line, where the two types grow at equal rates.

 
\par

We now consider what happens when two populations of cell types can exchange metabolites (i.e.\ $D>0$). For any initial mixed population, $n_1, n_2 > 0$, there is a Nash equilibrium choice of production rates where neither organism can improve its growth rate by changing its production. The growth rates of each cell type at this Nash equilibrium are not necessarily the same. If the growth rates are different, then one cell type will increase in relative frequency. This will alter the relative frequencies of the two cell types $n_1, n_2$ and could lead cell types to adapt to the new frequencies by changing their production. This process continues until either the growth rates of the two cell types are equal and their relative frequencies are stable or one cell type is driven towards extinction (zero frequency). For our choice of growth functions, production constraints, and parameters, there is always a unique stable equilibrium $n_1^*$ in terms of the relative frequencies of cell types. This means that for a given set of metabolic efficiencies all mixed populations will approach the same equilibrium values of relative frequency. Of course, a change in the efficiency of producing a metabolite could alter this equilibrium.
\par


\begin{figure}
    \begin{center}
    \includegraphics[width=0.99\linewidth]{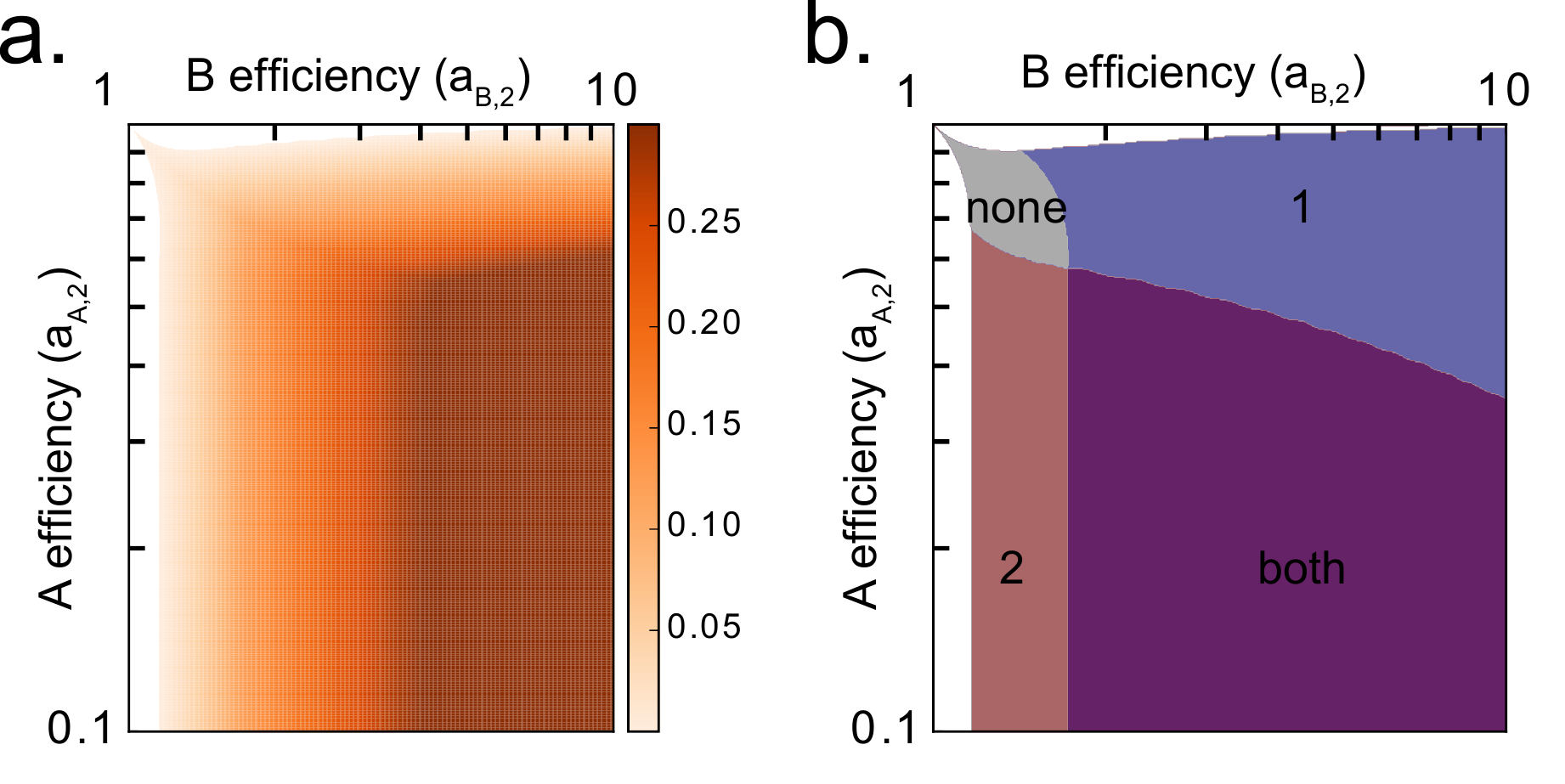} 
    \end{center}
    \caption{{\bf Specialization and the benefit of trade.} \textbf{a)}  The equilibrium growth rate of a population of two coexisting cell types is larger than what either would be able to achieve alone. Here, we plot the difference between the growth rate of a population of two cell types (one with the efficiencies shown and a reference cell type with $a_{A,1}=a_{B,1}$) and that of the surviving cell type in the absence of diffusion. All areas of coexistence in the two cell type population grow faster than the clonal cell population. \textbf{b)} The higher growth rate is achieved by cell types shifting production towards the metabolite in which they have higher efficiency than the other cell type. This shift may be complete for both types (purple), only for cell type 1 (blue) or 2 (red), or for neither (gray). \label{fig:together}}
\end{figure}


We compute the equilibrium $n_1^*$ as a function of the relative efficiencies of producing metabolites $A$ and $B$. As before, we hold one cell type, $i=1$, fixed in terms of its efficiencies and vary the efficiencies for the other cell type, $i=2$. When one cell type is better than the other at both production tasks, the only stable equilibrium is that its fraction of the population approaches one, and the other cell type is driven to extinction (results not shown). This trivial result seemingly contradicts the notion of comparative advantage, familiar from Ricardian economics \cite{ricardo1821}, where there is a benefit from trade even if one agent is better at producing all goods. In fact, at a fixed value of the relative frequency $0<n_1<1$, comparative advantage does play a role in setting the Nash equilibrium, and both cell type populations benefit from the diffusive exchange. However, because cells can reproduce, if one cell type, say 1, is better than the other at both production tasks, it will always grow faster, and any relative frequency except $n_1=1$ will not be sustainable. For the rest of the paper, we ignore this case and consider instead the case where each cell type is more efficient than the other at producing one of the two metabolites.

In Figure \ref{fig:alone}b, the gray region indicates where a coexistence equilibrium is observed. This region is much expanded in comparison to the line in Figure \ref{fig:alone}a, and many more combinations of efficiencies lead to coexistence. Even though neither cell type is more efficient than the other in the production of both metabolites, there is still a region of parameter space in which there is failure of coexistence. For example, when cells of type 2 are significantly worse at producing $A$ than their counterparts but only marginally better at producing $B$, then the system tends toward an equilibrium where cell type 2 goes extinct ($n_2 \to 0$). Similarly, there is a corresponding region where cells of type 2 are significantly better at producing $B$ but only marginally worse at producing $A$, and they take over the population ($n_2 \to 1$). 

Where coexistence occurs, we find that the growth rate (equal for the two cell types, by definition of the equilibrium) is larger than either cell type would have been able to achieve in isolation (see Figure \ref{fig:together}a). By concentrating production to the metabolite each cell type is better at producing, both cell types experience an increased growth rate. This result has been found in other, different models of microbial trade \cite{Tasoff:2015id,Enyeart:2015cl}. In our model, the advantage of trade is achieved even when specialization is not complete, i.e.\ when a cell type produces both metabolites. In Figure \ref{fig:together}b, we show the regions in parameter space where either both, one, or neither of the cell types specialize completely. In general, the highest growth rates occur where both cell types completely specialize, though there are regions of high growth where only cell type 1 completely specializes. In all cases, the increased growth rate resulting from trade, i.e. compared to growth in isolation, does not require any global coordination between the cell types. Rather, it emerges from each cell type producing what maximizes its own growth rate.

Until now, we have investigated primarily what conditions permit coexistence. However, if we consider also the resulting population composition and growth rates, we find that the interplay between the three types of dynamic variables in our model can lead to seemingly paradoxical phenomena. We illustrate three salient examples below.

\subsection*{Paradox 1: the curse of increased efficiency}
\begin{figure}
    \begin{center}
	\includegraphics[width=0.95\linewidth]{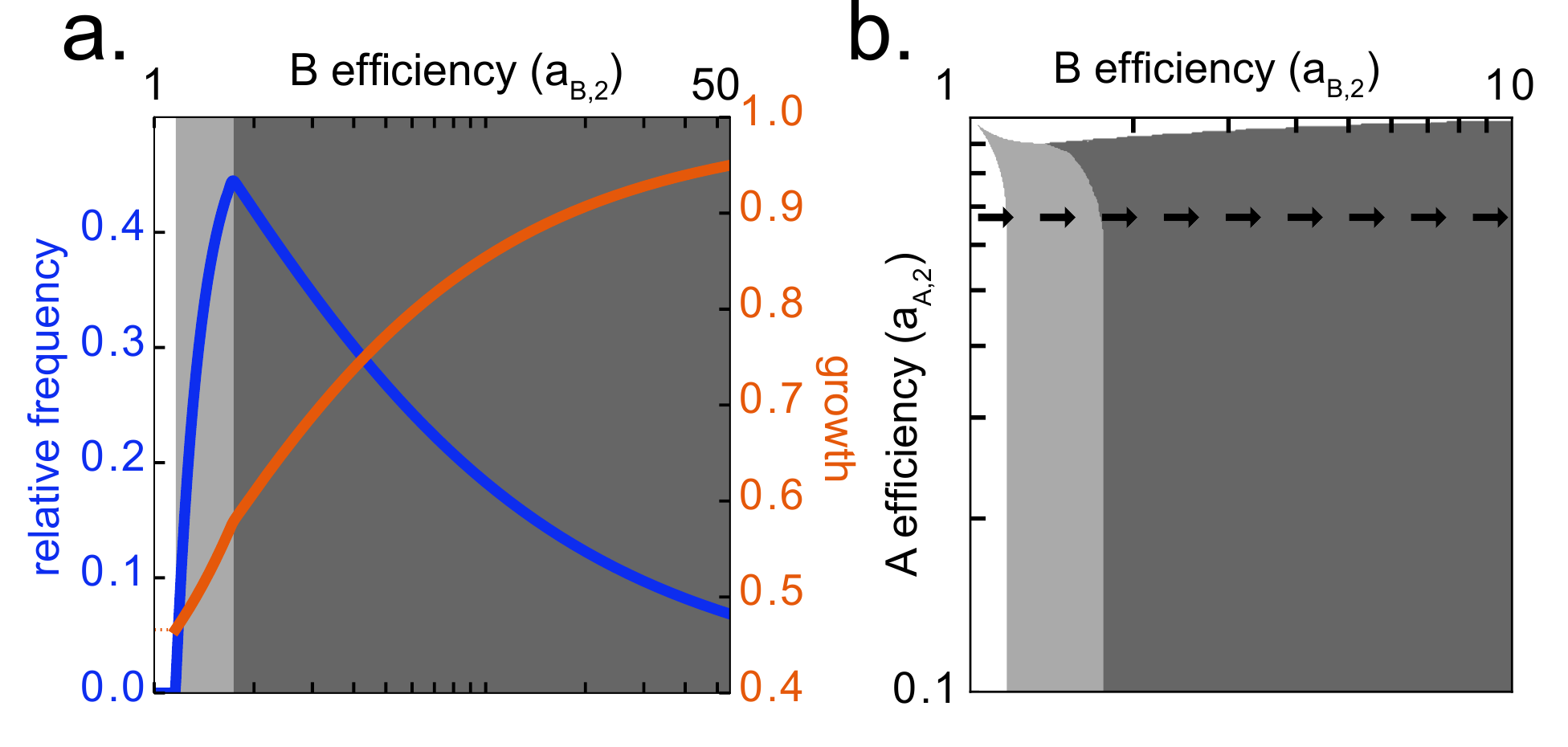}
    \end{center}
    \caption{\textbf{The curse of increased efficiency.} \textbf{a)} The relative frequency of type 2 cells (blue) and growth rate of the total population (orange) are shown as a function of the efficiency of type 2 cells in producing metabolite $B$. We fix $a_{A,2}=0.67$. As the efficiency of type 2 cells increases their relative frequency ultimately decreases. The population growth rate, however, increases with higher metabolic efficiency. 
	\textbf{b)} The shaded regions indicate where type 2 cell populations are increasing (light gray) or decreasing (dark gray) in relative frequency as they improve in efficiency in producing metabolite $B$, i.e. moving from left to right along the indicated line. In the dark gray region, the relative frequency of cell type 2 decreases towards 0 as its efficiency increases towards infinity.
	\label{fig:paradox1}}
\end{figure}

The first paradox concerns the relative frequency of cells of type 2 as a function of their metabolite production efficiency. Specifically, we consider a horizontal cross section of the parameter space in Figure \ref{fig:together}b where $a_{A,2}$ is fixed and $a_{B,2}$ varies. As cells of type 2 become better and better at producing $B$, their relative frequency at first increases as might be expected due to their increased productivity. However, at some point their relative frequency reaches a maximum and declines (see Figure \ref{fig:paradox1}a). Thus, even though the type 2 cells can produce more of the $B$ metabolite without decreasing production of $A$, they represent a smaller fraction of the population. This effect intensifies as the production efficiency of $B$ increases toward infinity, driving the fraction of type 2 cells in the population towards 0. 
\par
The region where this ``curse'' is in effect is illustrated in Figure \ref{fig:paradox1}b. It occurs in the area where cell type 1 is completely specialized and only makes the $A$ metabolite. Since cells need both $A$ and $B$ to grow, cell type 1 specializes in $A$ because there is enough $B$ in the environment provided by cell type 2 for it to forego production of $B$. Along a horizontal cross section, following the arrows in Figure \ref{fig:paradox1}b, cell type 2 gets more efficient at producing $B$ but cell type 1 has the same production capacity for $A$. Because type 1 cells are not getting better at producing $A$, their fraction of the population---needed to support the continued growth of both populations---increases. This phenomenon can be seen analytically in the limit of small degradation rate $\mu$ and under the assumption of full specialization, where $p_{B,1}=p_{A,2}=0$. In this case, we have $n_1 p_{A,1} = n_2 p_{B,2}$, and so $n_2 = a_{A,1}/(a_{A,1}+a_{B,2})$, where the fraction of type 2 cells is inversely proportionate to their efficiency in producing $B$.


\begin{figure}
    \begin{center}
    \includegraphics[width=0.95\linewidth]{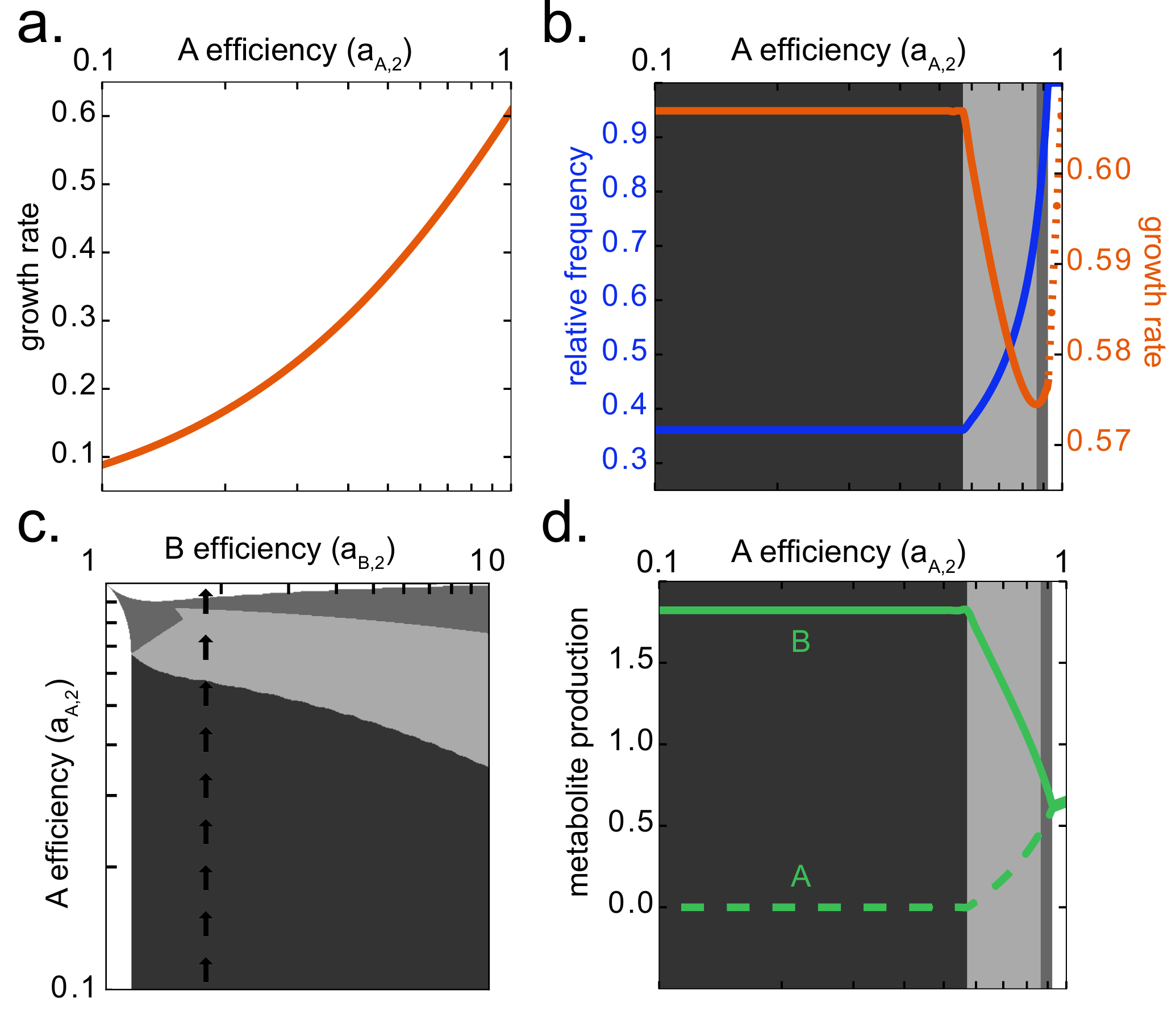}
    \end{center}
    \caption{\textbf{The curse of decreased inefficiency.} \textbf{a)} The growth rate of type 2 cells in isolation increases as they improve in efficiency at making $A$, the metabolite they produces more poorly ($a_{B,2}=1.82$, $a_{A,2}<1$). \textbf{b)} Similar to the previous plot, but in the presence of a reference cell type ($a_{B,1}=a_{A,2}=1$). Here, as the type 2 cells increase in efficiency at producing $A$, the population growth rate (orange) decreases in the light gray region. This also corresponds to an increase in frequency of the type 2 cell (blue). \textbf{c)} The shaded regions indicate where the population growth rate is constant (dark gray), decreasing (light gray), or increasing (middle gray) as type 2 cells improve in efficiency in producing metabolite $A$, i.e. moving up along the indicated line. \textbf{d)} The amount of $B$ and $A$ metabolite produced by type 2 cells is shown as a function of the efficiency in producing $A$. In the light gray region, where the population growth rate decreases, the type 2 cells shift production from the $B$ metabolite to the $A$ metabolite.
	\label{fig:paradox2}}
\end{figure}


Though the relative frequency of type 2 cells increases and decreases as $a_{B,2}$ increases, the population growth rate is always increasing. Therefore, even when the relative frequency of cell type 2 is decreasing, its total numbers might not be decreasing, because it is growing at a faster rate than it would be otherwise. The local effects---decreased relative frequency of cell type 2---are paradoxical, but the global effects---the population growth rate---are consistent with expectations. This observation partially resolves the paradox wherein a cell type population appears to suffer as a result of a gain in efficiency. However, in the next paradox, we will show that even when considering the growth rate instead of the relative frequency, a population can suffer as the result of a gain in efficiency.
\par

\subsection*{Paradox 2: the curse of decreased inefficiency}
The second paradox concerns the population growth rate of both cell types as a function of metabolite production efficiency. Here, we consider a vertical cross-section of Figure \ref{fig:together}b, where type 2 cells have a fixed efficiency of producing $B$, but a varying efficiency of producing $A$. Traversing up a vertical cross-section corresponds to type 2 cells being able to produce more $A$ but still not as much as type 1 cells. In a homogeneous population with only one cell type, any improved efficiency in production would correspond to an increased growth rate (see Figure \ref{fig:paradox2}a). However, in a mixed population, Figure \ref{fig:paradox2}b shows that as cells of type 2 get more efficient at producing the $A$ metabolite, the population growth rate decreases (before ultimately increasing). Thus, despite increased capacity to produce metabolites, the population grows more slowly. 

To explain this paradox, we consider the absolute maximum growth rate a population of two cell types could achieve assuming that they perfectly coordinated their production of metabolites. For this maximum growth rate to be sustainable, the growth rates of both cell types must be equal (or one cell type must have zero relative frequency); otherwise, the relative frequency of the cell types will change and the population will no longer be able to sustain this growth rate. There is no reason for the maximum sustainable growth rate to be achieved as a Nash equilibrium, and in general it is not. We determine the parameters $n_1, p_{A,1}, p_{B,1}, p_{A,2},$ and $p_{B,2}$ that correspond to the maximum sustainable growth rate and find that in almost all cases, this optimum is achieved when cell types fully specialize in their production of metabolites, i.e. $p_{B,1}=p_{A,2}=0$ and $p_{B,2},p_{A,1}>0$. The reason, then, that increasing the efficiency of cell type 2 to produce $A$ decreases the growth rate of the population is that it moves the Nash equilibrium away from complete specialization, that is, away from the steady state that achieves the maximum sustainable growth rate. This explains why the population growth rate starts decreasing at the same point that cell type 2 no longer specializes (see Figure \ref{fig:paradox2}c and Figure \ref{fig:paradox2}d). Interestingly there is a small range for $a_{B,2}$ for which there are two cycles of decreasing and increasing growth rate, corresponding to type 2 cells shifting production and then type 1 cells shifting production (see Figure \ref{fig:doublehump}).

\par

\begin{figure}
    \begin{center}
	\includegraphics[width=0.9\linewidth]{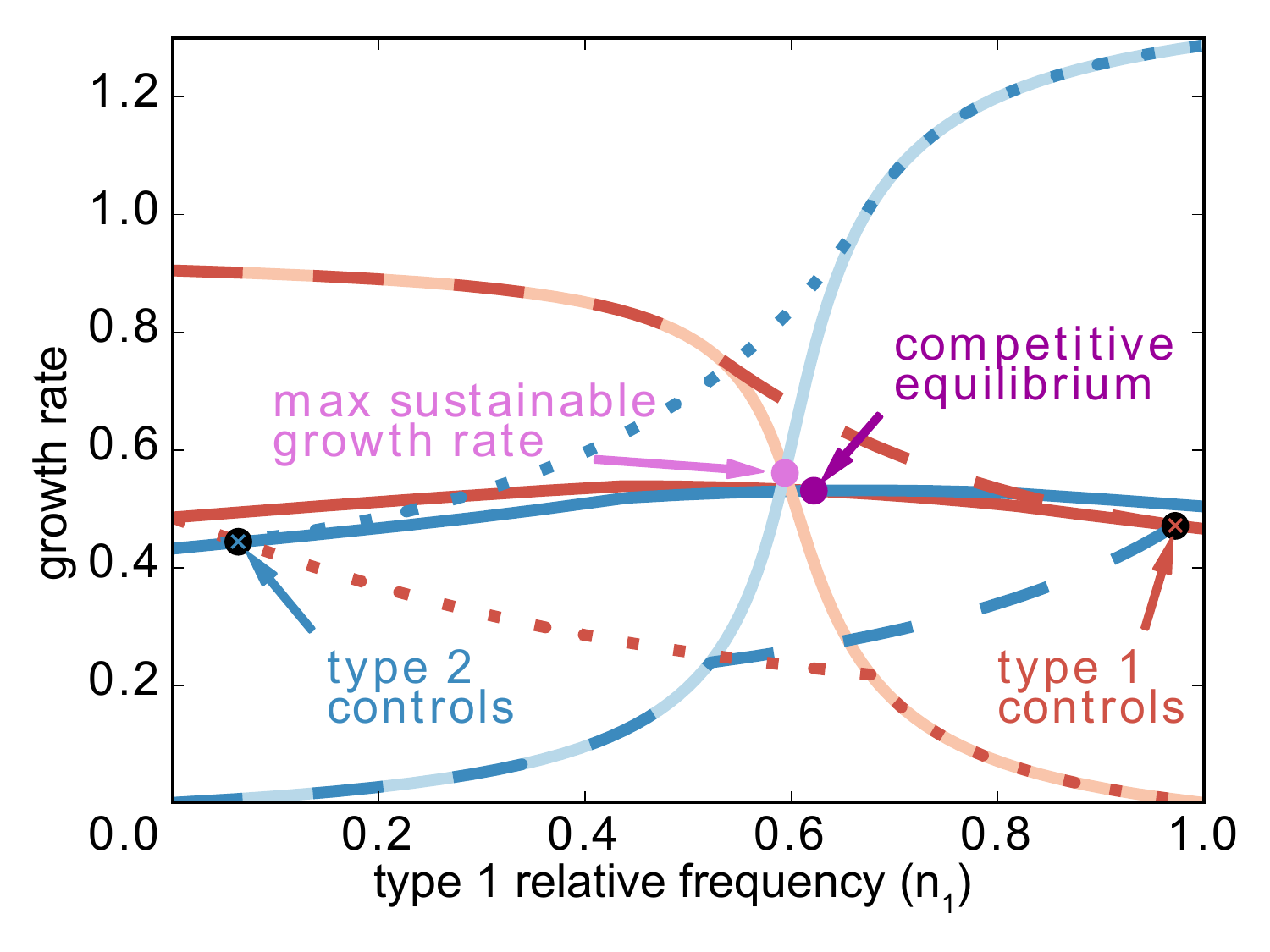}
    \end{center}
    \caption{ {\bf Different types of equilibria and the growth cost of control.} The steady-state growth rate of cells of type 1 (red) and type 2 (blue) under four different scenarios:
    each type maximizing its own growth (dark solid lines), both types maximizing the growth of cells of type 1 (dashed), both types maximizing the growth
    of cells of type 2 (dotted), and perfect coordination, where each cell type produces only a single metabolite (light solid lines).
    The production efficiencies used in this example are $a_{A,1}=a_{B,1}=1$, $a_{A,2}=0.67$, and $a_{B,2}=1.49$. The resulting
    population dynamic equilibria are marked. The growth rate of the competitive equilibrium is closer to the maximum sustainable growth rate than either equilibria reached when a single cell type is in control.
    \label{fig:paradox3}}
\end{figure}


\subsection*{Paradox 3: the curse of control}
The final paradox focuses on the population growth rate as a function of how metabolic production is determined. Until now, we have assumed that the two cell types are both choosing their own production so as to maximize their own growth rate. Here, we consider what happens if one cell type is able to determine both its own production rates as well as those of the other cell type. This situation occurs in some game theoretic settings \cite{FolkTheorem}, where a single player can force others to follow a particular strategy of their choice, for example by playing a punitive strategy when the other players deviate. In the case of microbes, we imagine that a microbial population has evolved the ability to manipulate its partner. 

We implement the manipulation by assuming that the production rates for both cell types are chosen to maximize the growth rate of type 1 cells, regardless of the resulting growth rate for cells of type 2. We repeat the numerical process as before where we solve for the steady-state growth rates at a given value of $n_1$, and depending on the relative value of growth rates, either increase or decrease $n_1$. Since cell type 1 is controlling production, it continually increases in relative frequency until the growth rates of the two cell types are identical. We find that while the resulting equilibrium has a larger proportion of cells of type 1 compared with the situation achieved by the competitive Nash equilibrium, the population grows at a slower rate.

To understand why this paradox occurs, we choose a set of production efficiencies and compute the steady-state growth rates of the two cell types as a function of their relative frequencies.
Figure \ref{fig:paradox3} shows the growth rates under four scenarios: (1) each cell type maximizing its own growth rate, (2) both types maximizing growth rate of cells of type 1, (3) both type maximizing the growth rate of cells of type 2, and (4) complete specialization, where each cell type produces only a single metabolite. The population dynamic equilibrium is achieved when the two growth rates are equal. Though the competitive setup leads to a population dynamic equilibrium with a growth rate that is not as large as the maximum sustainable growth rate, it comes closer to this optimum than do the equilibria that result from maximizing the growth of either single cell type. 

This paradox demonstrates that if a cell type controls another so as to maximize its immediate growth rate, then it effectively sacrifices its long-term growth rate. This implies that there is always some long enough time horizon for 
which this tradeoff will result in fewer cells of the controller. 
Suppose that $dN_i/dt = \zeta g_i N_i$, where $\zeta$ determines the typical population dynamics time scale relative
to the typical times over which the chemical concentrations evolve. Then, we can determine how long it takes
for the number of type 1 cells in a population following the competitive production dynamics to overtake the
number of type 1 cells in a population following the production dynamics controlled by type 1 cells. If the
two cell types start at equal frequency, this will happen after a time $t=13.5\zeta^{-1}$, at which point
the number of type 1 cells in both populations would have grown by a factor of $1420$ (see Figure \ref{fig:gtraj}).

\section*{Discussion}
Microbes constantly face decisions about which metabolites to produce. These decisions depend on what metabolites are present in the environment, which, in turn, can be affected by the abundance and production decisions of other microbes. Here we introduce a simple, general mathematical model to understand the interplay between microbial production decisions and population dynamics. Using this model, we identify the conditions that permit coexistence among different species and discover three paradoxical behaviors that demonstrate the unusual feedback between individual-level production and population-level dynamics. 

In our model, beneficial trade emerges naturally as metabolites diffuse in and out of cells as each organism maximizes its own growth rate. We find that different microbes are able to coexist across a broad range of production costs/efficiencies, and in all cases of coexistence, microbes grow faster than if they were isolated from one another. Coexistence occurs only when the species are each more efficient at producing a different resource. However, this is not a sufficient condition. If one organism is much more efficient at producing one resource and only marginally worse at producing the other, then it can drive the other species to zero frequency. Thus, there is some threshold for production efficiencies that permits coexistence. In our model, this threshold depends on the growth and production functions of each microbe as well as the diffusion and metabolite consumption rates. While we investigated the simple case in which each microbe has the same growth function and similar production constraints, in real biological systems it is likely that these may differ across species, and these differences may result in a richer and more complex set of dynamics \cite{Hoeksema:2003du}.

Another consequence of our model is the natural emergence of a division of labor. At each iteration of our model, each microbial species made a production decision that maximized its growth rate in the current environment. Through this simple process, we observed that each microbe shifted its production to the metabolite it is better at producing. Although the microbes did not become complete obligates, we found that this increased specialization led to a higher population growth rate without any external coordination. We note that for some fixed values of production efficiencies/costs, the population as a whole could grow fastest if each microbe completely specialized and this outcome was a stable Nash equilibrium. 

The curse of decreased inefficiency provides a mechanism by which the division of labor in mutual obligates (the {\it mutualism} category in Table \ref{table-models}) evolves without built-in returns to scale or benefit of specialization in the production efficiency. In some models of microbial exchange, when one organism loses the ability to make a resource it grows faster due to a built-in benefit of specialization \cite{Estrela:2016kn}. This leads to a rapid loss of functionality in co-evolving species such that they become mutually reliant on one another. In turn, this loss of functionality can lead to the situation featured in the Black Queen hypothesis discussed earlier \cite{Morris:2012bd}. Our model shows that such increased growth does not require any built-in benefits of specialization. It can simply emerge as a consequence of the fact that a loss in efficiency forces one species to bring its production strategy closer to the globally optimal situation of complete specialization, i.e.\ the inverse of paradox 2. Figure \ref{fig:together}a shows that the population can grow faster if type 2 cells either increase their efficiency in producing $B$ or decrease their efficiency in producing $A$.

Another paradox we uncovered is the curse of increased efficiency, in which as one species becomes more efficient at producing a resource, it becomes rarer in the population. Note that a species that produced both resources less efficiently would experience a similar decline in population.  Although these two scenarios exhibit similar qualitative behavior there are important differences in population structure and stability. In the case of the more efficient species, even though it is rare, it is significant by virtue of its metabolic contribution. If it went extinct due to some stochastic fluctuation, then the population growth rate would sharply decrease because of the dependency of the more abundant species. In the case of the less efficient species, the more abundant species does not have any such dependency and would experience little change in its growth rate if the less efficient species went extinct. These two scenarios may happen in real populations and without a detailed understanding of the interdependencies that evolved through trade we may make incorrect inferences about these systems. Indeed, in larger microbial consortia, our findings suggest that low abundance organisms may not necessarily be less fit and, potentially, they could be essential to the fast growth rate of the community.

The third paradox we uncovered, the curse of control, shows how exploitation of one species by another can lead the whole system to a worse growth rate. This paradox demonstrates a problem with forms of parasitism and cheating via chemical manipulation \cite{Keller:2006hq}. The short-term gains in population abundance that arise by these strategies might lead to long-term losses in the population growth rate. Depending on the length of time of an association, it may be more beneficial to compete with another microbe than to exploit it.

In this paper we restricted ourselves to the study of a particularly simple case of a much more general model. Though we consider only two metabolites and two cell types with equal requirements for growth, we not only illustrate the uncoordinated emergence of beneficial trade within a coexisting population, but also uncover a rich landscape of unexpected outcomes. Our model can be generalized to study the interaction of any number of cell types, exchanging any number of valuable molecular species, with arbitrary growth rates given as functions of the concentrations of these molecules, and with arbitrary constraints on their production rates. We expect that by adding more complexity to our model we will be able to model a large range of emergent behavior that may be present in real microbial community and may run counter to common intuition and implicit assumptions about the driving principles in these communities.

\textbf{Acknowledgments.} We acknowledge helpful discussions with David Wolpert. Y.\ K.\ and E.\ L.\ acknowledge the support of the Santa Fe Institute through the Omidyar Fellowship.

\bibliography{Metab}

\begin{thebibliography}{10}

\bibitem{Nadell:2016bl}
Carey~D. Nadell, Knut Drescher, and Kevin~R. Foster.
\newblock {Spatial structure, cooperation and competition in biofilms}.
\newblock {\em Nature Reviews Microbiology}, 14(9):589--600, 2016.

\bibitem{Sachs:2012hn}
J.~L. Sachs and A.~C. Hollowell.
\newblock {The Origins of Cooperative Bacterial Communities}.
\newblock {\em mBio}, 3(3):e00099--12--e00099--12, 2012.

\bibitem{Kouzuma:2015fj}
Atsushi Kouzuma, Souichiro Kato, and Kazuya Watanabe.
\newblock {Microbial interspecies interactions: recent findings in syntrophic
  consortia}.
\newblock {\em Frontiers in Microbiology}, 6:477, 2015.

\bibitem{Johns:2016kc}
Nathan~I. Johns, Tomasz Blazejewski, Antonio~LC Gomes, and Harris~H. Wang.
\newblock {ScienceDirect Principles for designing synthetic microbial
  communities}.
\newblock {\em Current Opinion in Microbiology}, 31:146--153, 2016.

\bibitem{Werner:2014dy}
G.~D.~A. Werner, J.~E. Strassmann, A.~B.~F. Ivens, D.~J.~P. Engelmoer,
  E.~Verbruggen, D.~C. Queller, R.~Noe, N.~C. Johnson, P.~Hammerstein, and
  E.~T. Kiers.
\newblock {Evolution of microbial markets}.
\newblock {\em Proceedings of the National Academy of Sciences of the United
  States of America}, 111(4):1237--1244, 2014.

\bibitem{Tasoff:2015id}
Joshua Tasoff, Michael~T. Mee, and Harris~H. Wang.
\newblock {An Economic Framework of Microbial Trade}.
\newblock {\em PloS one}, 10(7):e0132907--20, 2015.

\bibitem{Hammerstein:2016bn}
Peter Hammerstein and Ronald No{\"e}.
\newblock {Biological trade and markets.}
\newblock {\em Philosophical Transactions of the Royal Society B: Biological
  Sciences}, 371(1687):20150101--12, 2016.

\bibitem{Morris:2015dq}
J.~Jeffrey Morris.
\newblock {Black Queen evolution: the role of leakiness in structuring
  microbial communities}.
\newblock {\em Trends in Genetics: TIG}, 31(8):475--482, 2015.

\bibitem{Pande:2015hb}
Samay Pande, Shraddha Shitut, Lisa Freund, Martin Westermann, Felix Bertels,
  Claudia Colesie, Ilka~B. Bischofs, and Christian Kost.
\newblock {Metabolic cross-feeding via intercellular nanotubes among bacteria}.
\newblock {\em Nature Communications}, 6:6238, 2015.

\bibitem{Estrela:2016kn}
Sylvie Estrela, J.~Jeffrey Morris, and Benjamin Kerr.
\newblock {Private benefits and metabolic conflicts shape the emergence of
  microbial interdependencies.}
\newblock {\em Environmental microbiology}, 18(5):1415--1427, 2016.

\bibitem{Cordero:2012bi}
Otto~X. Cordero, Laure-Anne Ventouras, Edward~F. DeLong, and Martin~F. Polz.
\newblock {Public good dynamics drive evolution of iron acquisition strategies
  in natural bacterioplankton populations}.
\newblock {\em Proceedings of the National Academy of Sciences of the United
  States of America}, 109(49):20059--20064, 2012.

\bibitem{Schink:2002uy}
Bernhard Schink.
\newblock {Synergistic interactions in the microbial world}.
\newblock {\em Antonie van Leeuwenhoek}, 81(1):257--261, 2002.

\bibitem{Morris:2013ja}
Brandon E.~L. Morris, Ruth Henneberger, Harald Huber, and Christine
  Moissl-Eichinger.
\newblock {Microbial syntrophy: interaction for the common good}.
\newblock {\em FEMS Microbiology Reviews}, 37(3):384--406, 2013.

\bibitem{Wyatt:2014jv}
Gregory A.~K. Wyatt, E.~Toby Kiers, Andy Gardner, and Stuart~A. West.
\newblock {A biological market analysis of the plant-mycorrhizal symbiosis}.
\newblock {\em Evolution}, 68(9):2603--2618, 2014.

\bibitem{Kummel:2006bl}
Miroslav Kummel and Stephen~W. Salant.
\newblock {The economics of mutualisms: optimal utilization of mycorrhizal
  mutualistic partners by plants}.
\newblock {\em Ecology}, 87(4):892--902, 2006.

\bibitem{McInerney:2008hp}
Michael~J. McInerney, Christopher~G. Struchtemeyer, Jessica Sieber, Housna
  Mouttaki, Alfons J.~M. Stams, Bernhard Schink, Lars Rohlin, and Robert~P.
  Gunsalus.
\newblock {Physiology, Ecology, Phylogeny, and Genomics of Microorganisms
  Capable of Syntrophic Metabolism}.
\newblock {\em Annals of the New York Academy of Sciences}, 1125(1):58--72,
  2008.

\bibitem{Bull:2009dr}
James~J. Bull and William~R. Harcombe.
\newblock {Population dynamics constrain the cooperative evolution of
  cross-feeding.}
\newblock {\em PloS one}, 4(1):e4115, 2009.

\bibitem{Eberhard:1975hi}
M.~J.~W. Eberhard.
\newblock {The Evolution of Social Behavior by Kin Selection}.
\newblock {\em Quarterly Review of Biology}, 50(1):1--33, 1975.

\bibitem{Sachs:2004fe}
Joel~L. Sachs, Ulrich~G. Mueller, Thomas~P. Wilcox, and James~J. Bull.
\newblock {The Evolution of Cooperation}.
\newblock {\em The Quarterly review of biology}, 79(2):135--160, 2004.

\bibitem{Doebeli:1998uo}
Michael Doebeli and Nancy Knowlton.
\newblock {The evolution of interspecific mutualisms}.
\newblock {\em Proceedings of the National Academy of Sciences of the United
  States of America}, 95(15):8676--8680, 1998.

\bibitem{Ghoul:2016cz}
M.~Ghoul and S.~Mitri.
\newblock {The Ecology and Evolution of Microbial Competition}.
\newblock {\em Trends in Microbiology}, 24(10):833--845, 2016.

\bibitem{Morris:2012bd}
J.~J. Morris, R.~E. Lenski, and E.~R. Zinser.
\newblock {The Black Queen Hypothesis: evolution of dependencies through
  adaptive gene loss}.
\newblock {\em mBio}, 3(2):e00036--12, 2012.

\bibitem{Biggs:2015gi}
Matthew~B. Biggs, Gregory~L. Medlock, Glynis~L. Kolling, and Jason~A. Papin.
\newblock {Metabolic network modeling of~microbial communities}.
\newblock {\em Wiley Interdisciplinary Reviews: Systems Biology and Medicine},
  7(5):317--334, 2015.

\bibitem{Stolyar:2007jh}
Sergey Stolyar, Steve Van~Dien, Kristina~Linnea Hillesland, Nicolas Pinel,
  Thomas~J. Lie, John~A. Leigh, and David~A. Stahl.
\newblock {Metabolic modeling of a mutualistic microbial community.}
\newblock {\em Molecular Systems Biology}, 3(1):92--14, 2007.

\bibitem{Harcombe:2014it}
William~R. Harcombe, William~J. Riehl, Ilija Dukovski, Brian~R. Granger, Alex
  Betts, Alex~H. Lang, Gracia Bonilla, Amrita Kar, Nicholas Leiby, Pankaj
  Mehta, Christopher~J. Marx, and Daniel Segre.
\newblock {Metabolic Resource Allocation in Individual Microbes Determines
  Ecosystem Interactions and Spatial Dynamics}.
\newblock {\em CellReports}, 7(4):1104--1115, 2014.

\bibitem{Klitgord:2010cs}
N.~Klitgord and D.~Segre.
\newblock {Environments that induce synthetic microbial ecosystems}.
\newblock {\em PLoS Computational Biology}, 6(11):e1001002, 2010.

\bibitem{Taillefumier2016}
T.~Taillefumier, A.~Posfai, Y.~Meir, and N.~S. Wingreen.
\newblock Bacterial cartels at steady supply.
\newblock 2016.
\newblock arXiv:1604.02733.

\bibitem{ricardo1821}
David Ricardo.
\newblock {\em On the Principles of Political Economy and Taxation}.
\newblock McMaster University Archive for the History of Economic Thought, 3
  edition, 1821.

\bibitem{Enyeart:2015cl}
Peter~J. Enyeart, Zachary~B. Simpson, and Andrew~D. Ellington.
\newblock {A microbial model of economic trading and comparative advantage}.
\newblock {\em Journal of theoretical biology}, 364:326--343, 2015.

\bibitem{FolkTheorem}
D.~Fudenberg, D.~Levine, and E.~Maskin.
\newblock The folk theorem with imperfect public information.
\newblock {\em Econometrica: Journal of the Econometric Society},
  62(5):997--1039, 1994.

\bibitem{Hoeksema:2003du}
Jason~D. Hoeksema and Mark~W. Schwartz.
\newblock {Expanding comparative-advantage biological market models:
  contingency of mutualism on partners' resource requirements and acquisition
  trade-offs.}
\newblock {\em Proceedings: Biological Sciences}, 270(1518):913--919, 2003.

\bibitem{Keller:2006hq}
Laurent Keller and Michael~G. Surette.
\newblock {Communication in bacteria: an ecological and evolutionary
  perspective}.
\newblock {\em Nature Reviews Microbiology}, 4:249--258, 2006.

\end{thebibliography}

\beginsupplement
\appendix
\section*{Supplementary material}
\subsection*{Effect of response speed}

In the main text, we use the simple assumption that cells' production decisions can be regulated on a much shorter time scale than the population dynamic time scale.
While this assumption is convenient, it is not necessary: even if the production decisions move toward their optimal value at a rate much slower than the population
growth rate, say through small and rare mutations, it is still the case that the only equilibrium situation is the same equilibrium we identify in the main text.

\begin{figure}
    \begin{center}
    \includegraphics[width=0.99\linewidth]{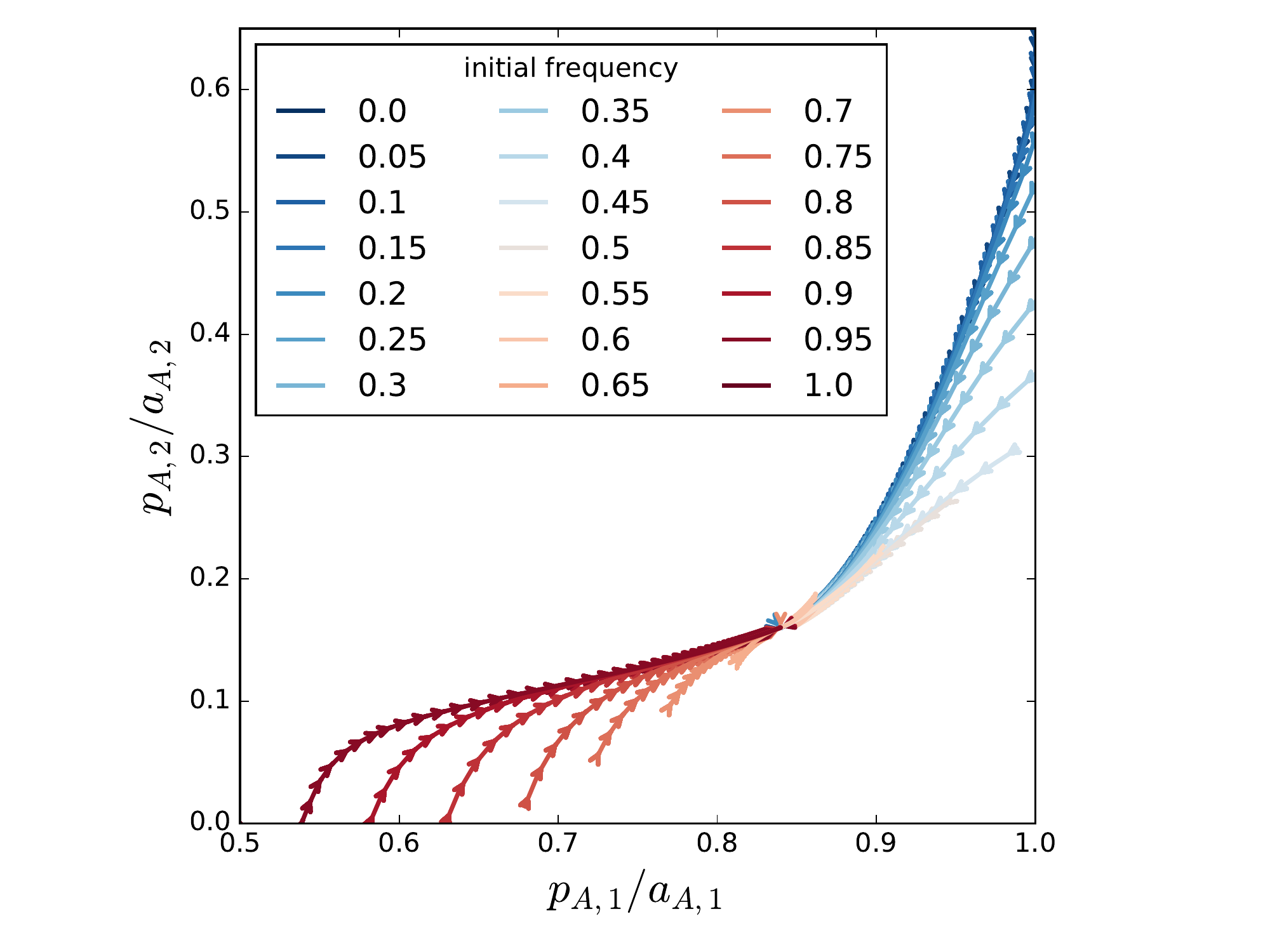} 
    \end{center}
    \caption{{\bf Effect of slower response.} Trajectories of solutions to Eq.\ \eqref{eq:SI_dyn} starting from the Nash equilibrium at some non-equilibrium relative
    frequency. The horizontal (vertical) axis shows the fraction of the budget devoted by cells of type 1 (2) to producing $A$. The production efficiencies used are $a_{A,1}=a_{B,1}=1$, $a_{A,2}=0.67$, and $a_{B,2}=1.49$, and
    the response speed parameter used is $\beta=0.05$. All trajectories arrive at the same equilibrium as found when the time scale for production decisions is much faster than population dynamics.
\label{fig:SI_dyn} }
\end{figure}

Consider the following dynamical system:
\begin{linenomath*}
\begin{equation}\label{eq:SI_dyn}
    \begin{aligned}
	\frac{dp_{X,i}}{dt} &= \beta(p_{X,i}^*-p_{X,i})\text{,} \qquad\text{for }i=1,2\text, \quad X=A,B\\
	\frac{dn_{1}}{dt} &= n_1(1-n_1)(g_1-g_2)\text,
    \end{aligned}
\end{equation}
\end{linenomath*}
where $p_{X,i}^*$ is the optimal production decision for cell type $i$ in the present environment, $g_i$ is the growth
rate of cell type $i$ under the present frequencies and production decision, and $\beta$ is a parameter determining the
relative speed at which production decisions adjust toward the optimum.
In Figure \ref{fig:SI_dyn}, we show trajectories of this system for $\beta=0.05$, starting at different initial relative
frequencies. Even with the lag between population dynamics and production decisions, we find that trajectories converge to the same equilibrium as found in our original analyses.

 
\subsection*{Effect of diffusion and degradation rates}

\begin{figure}
    \begin{center}
    \includegraphics[width=0.32\linewidth]{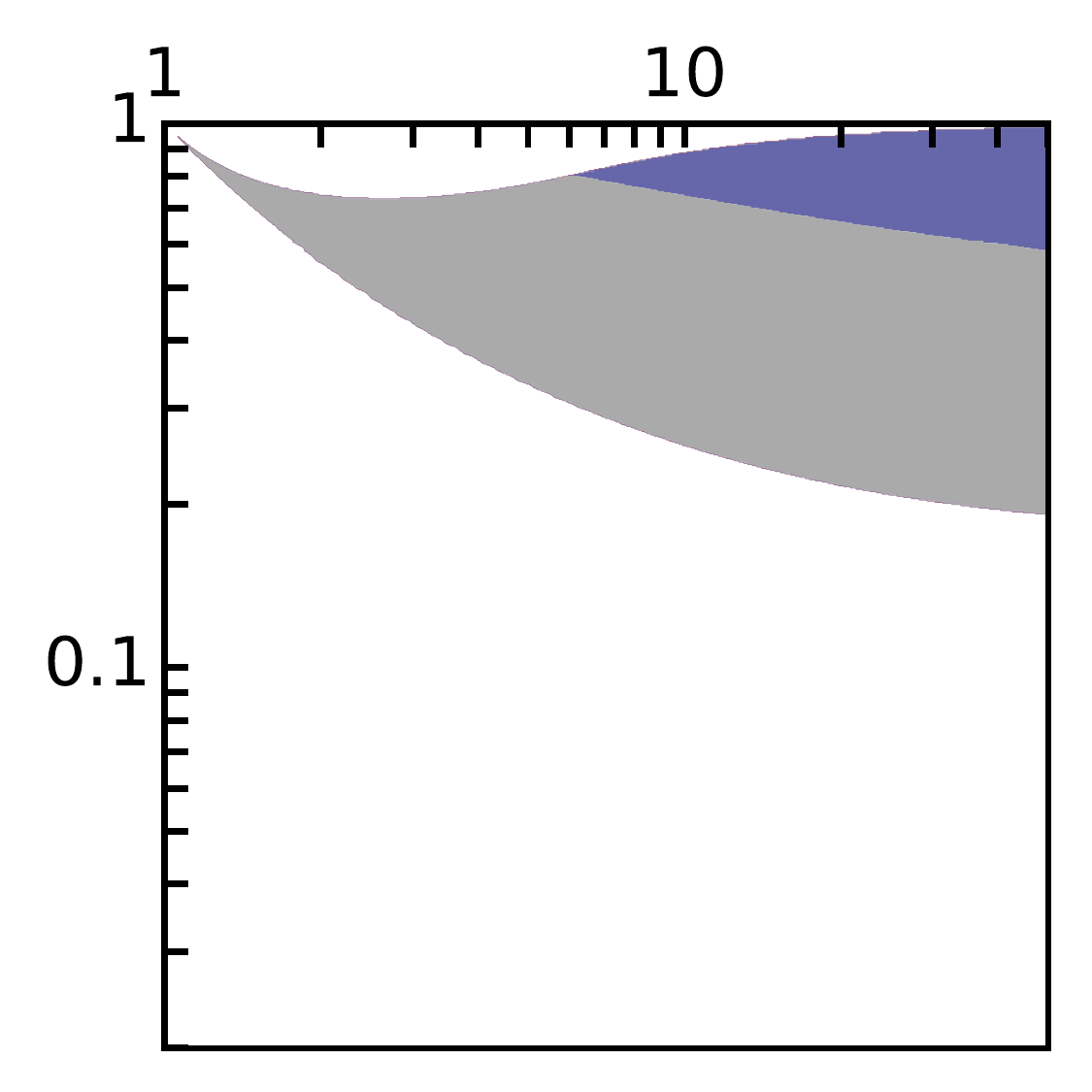} 
    \includegraphics[width=0.32\linewidth]{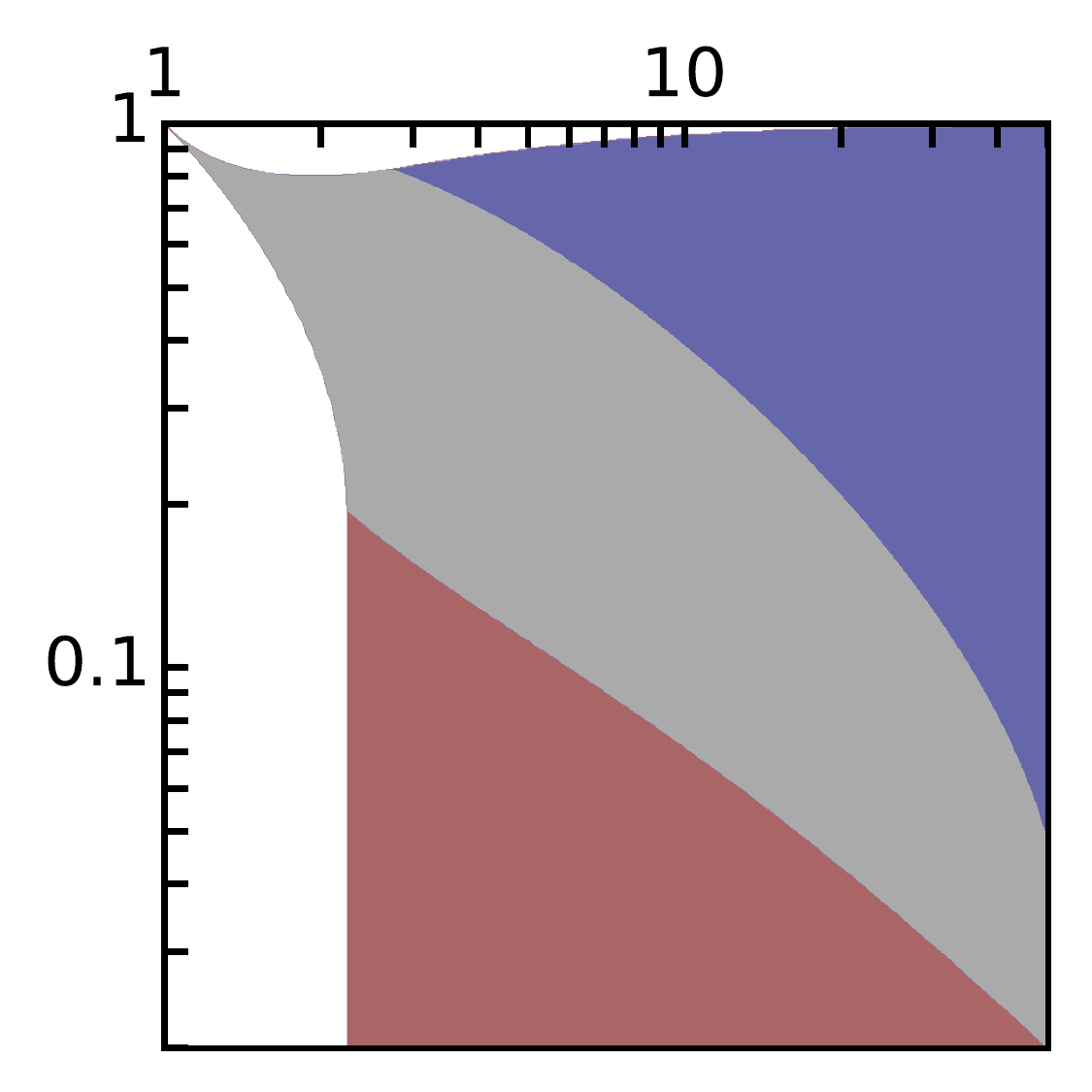} 
    \includegraphics[width=0.32\linewidth]{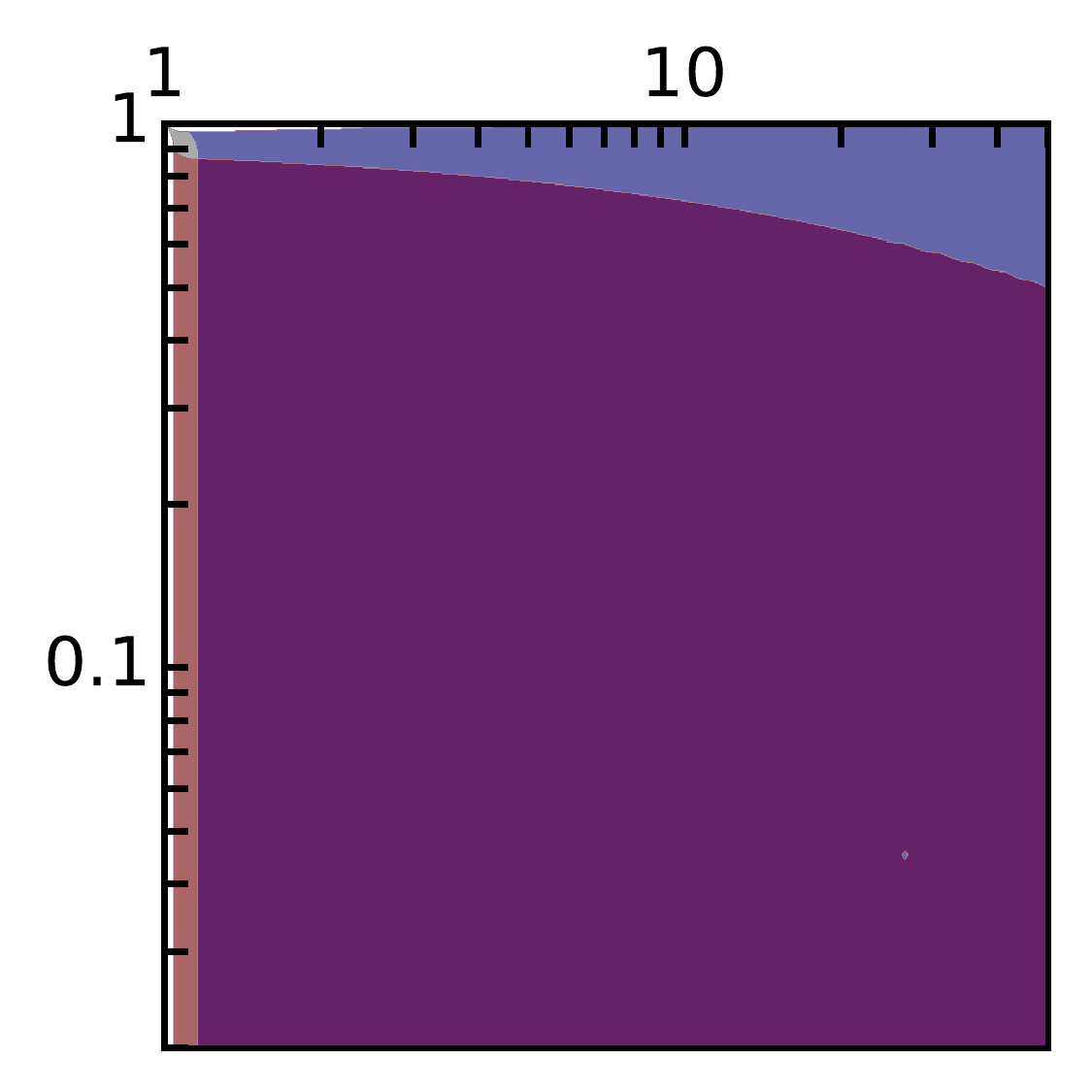} 
    \end{center}
    \caption{{\bf Effect of diffusion rate.} The region of parameter space allowing coexistence grows as a function
    of the diffusion rate $D$. The panels of this figure are analogs of Figure \ref{fig:together}b for $D=0.5$ (left), $D=1$ (center), and $D=10$ (right).
\label{fig:diffD} }
\end{figure}

\begin{figure}
    \begin{center}
    \includegraphics[width=0.32\linewidth]{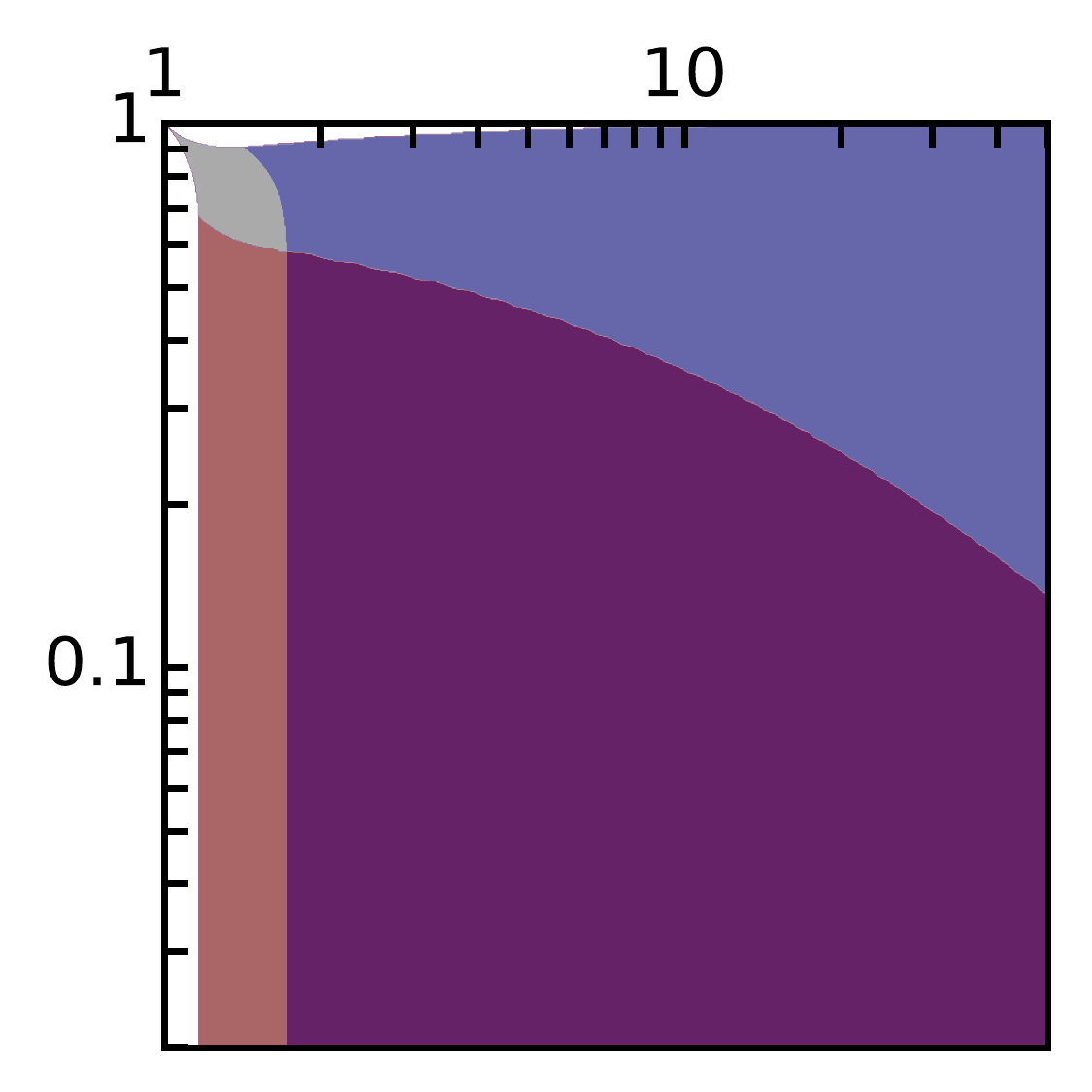} 
    \includegraphics[width=0.32\linewidth]{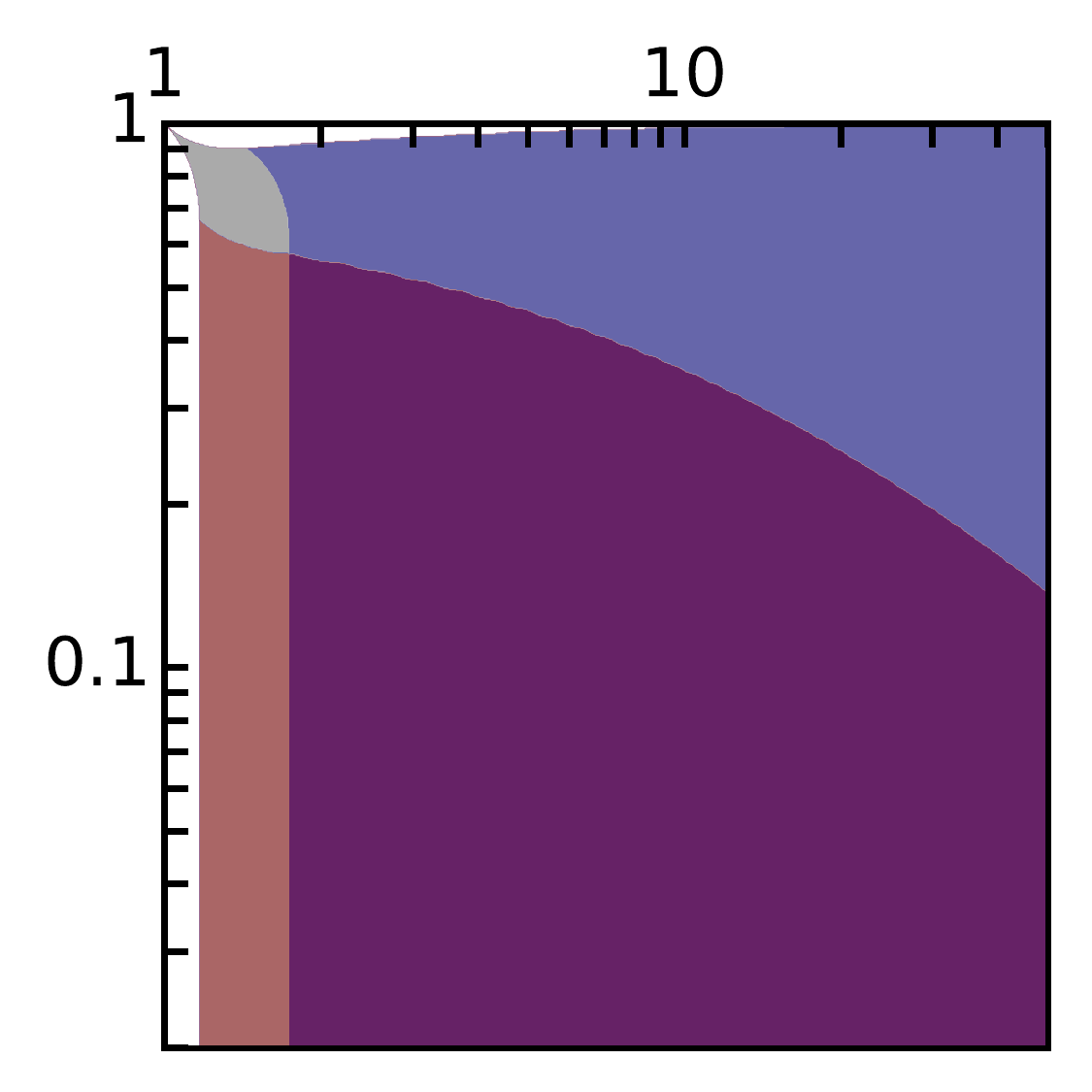} 
    \includegraphics[width=0.32\linewidth]{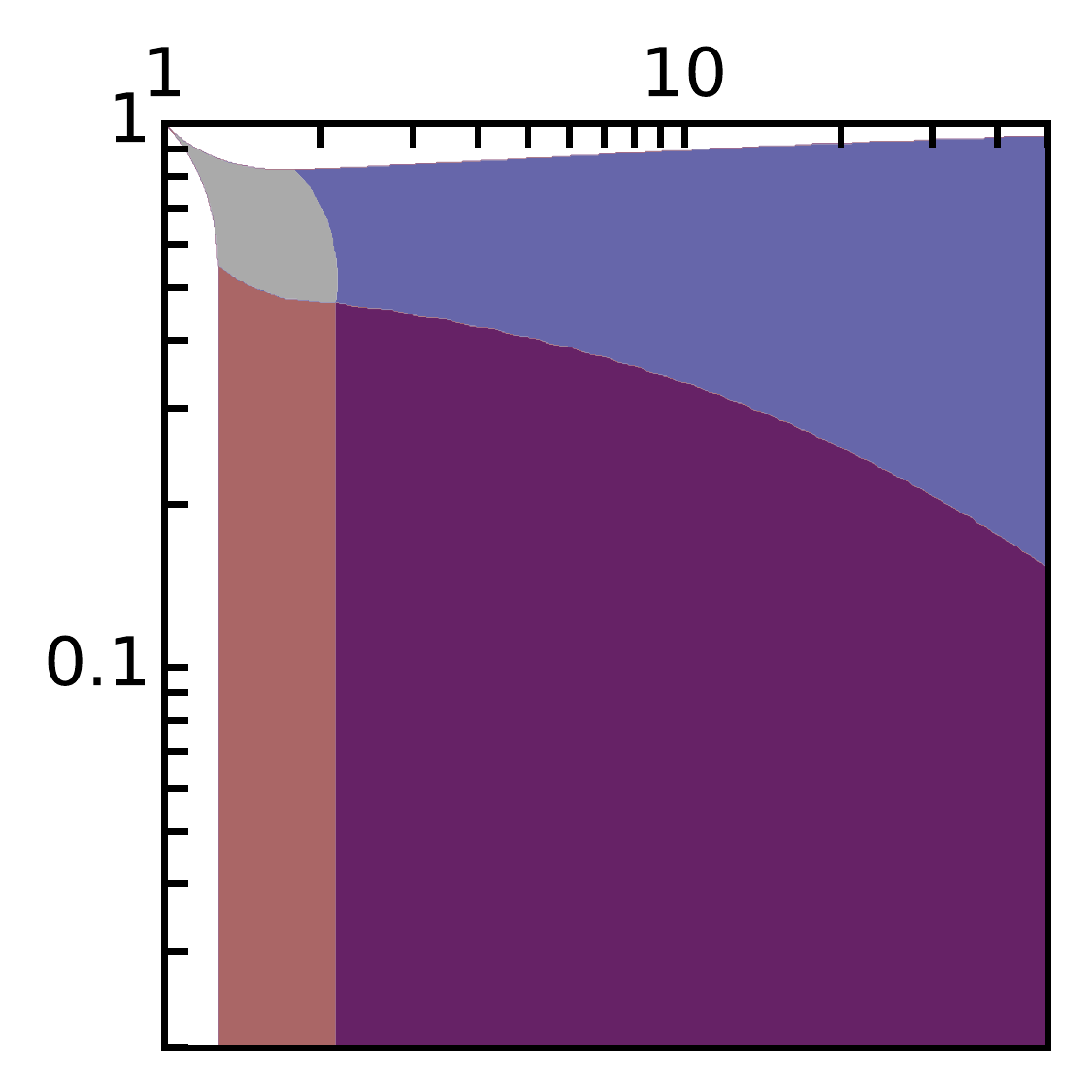} 
    \end{center}
    \caption{{\bf Effect of degradation rate.} The effect of the degradation rate $\mu$ is minimal. The panels of this figure are analogs of Figure \ref{fig:together}b for $\mu=0.02$ (left), $\mu=0.1$ (center), and $\mu=1$ (right).
\label{fig:diffmu} }
\end{figure}

In Figures \ref{fig:diffD} and \ref{fig:diffmu}, we illustrate the effects of varying the rates of diffusion of metabolites, $D$, and degradation of metabolites, $\mu$, relative to the rate at which they are consumed by the growth reaction. (We have taken the reaction rate constant to be $1$, setting the units of time). As we have shown in the main text, changing the diffusion coefficient from $D=0$ to $D=3$ expands the region of parameter space allowing coexistence from a 1-dimensional curve to a region occupying most of the quadrant where each cell type is more efficient in producing some metabolite. As we can see in Figure \ref{fig:diffD}, this change is gradual as a function of $D$, and as $D$ tends to infinity, the coexistence region expands to the entire quadrant. Moreover, the region of complete specialization by both cell types expands and also tends to extend to the entire quadrant.
By comparison, the changes that occur as a function of the degradation rates are minimal, but a smaller $\mu$ does
tend to support coexistence and complete specialization.

 
 \begin{figure}
    \begin{center}
    \includegraphics[width=0.6\linewidth]{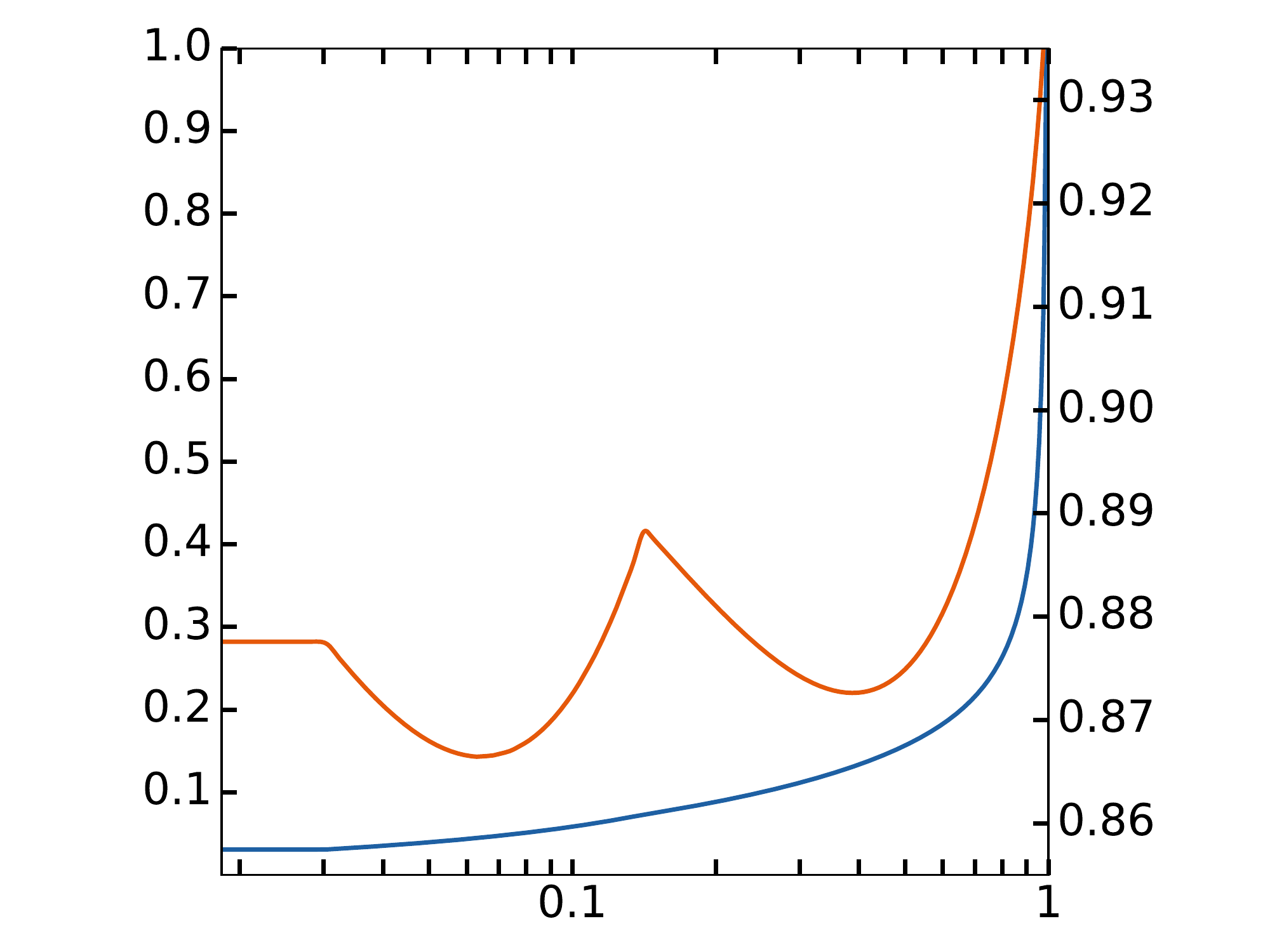} 
    \end{center}
    \caption{{\bf The curse of decreased inefficiency redux.} In some cases, our model predicts two noncontiguous regions of decreasing growth rate as a function of increased
    efficiency. Here we show the analog of Figure \ref{fig:paradox2}b for the case where $D=1$, $a_{A,1}=a_{B,1}=1$, $a_{B,2}=30$, and $a_{A,2}$ varies.
\label{fig:doublehump} }
\end{figure}



 \begin{figure}
    \begin{center}
    \includegraphics[width=0.6\linewidth]{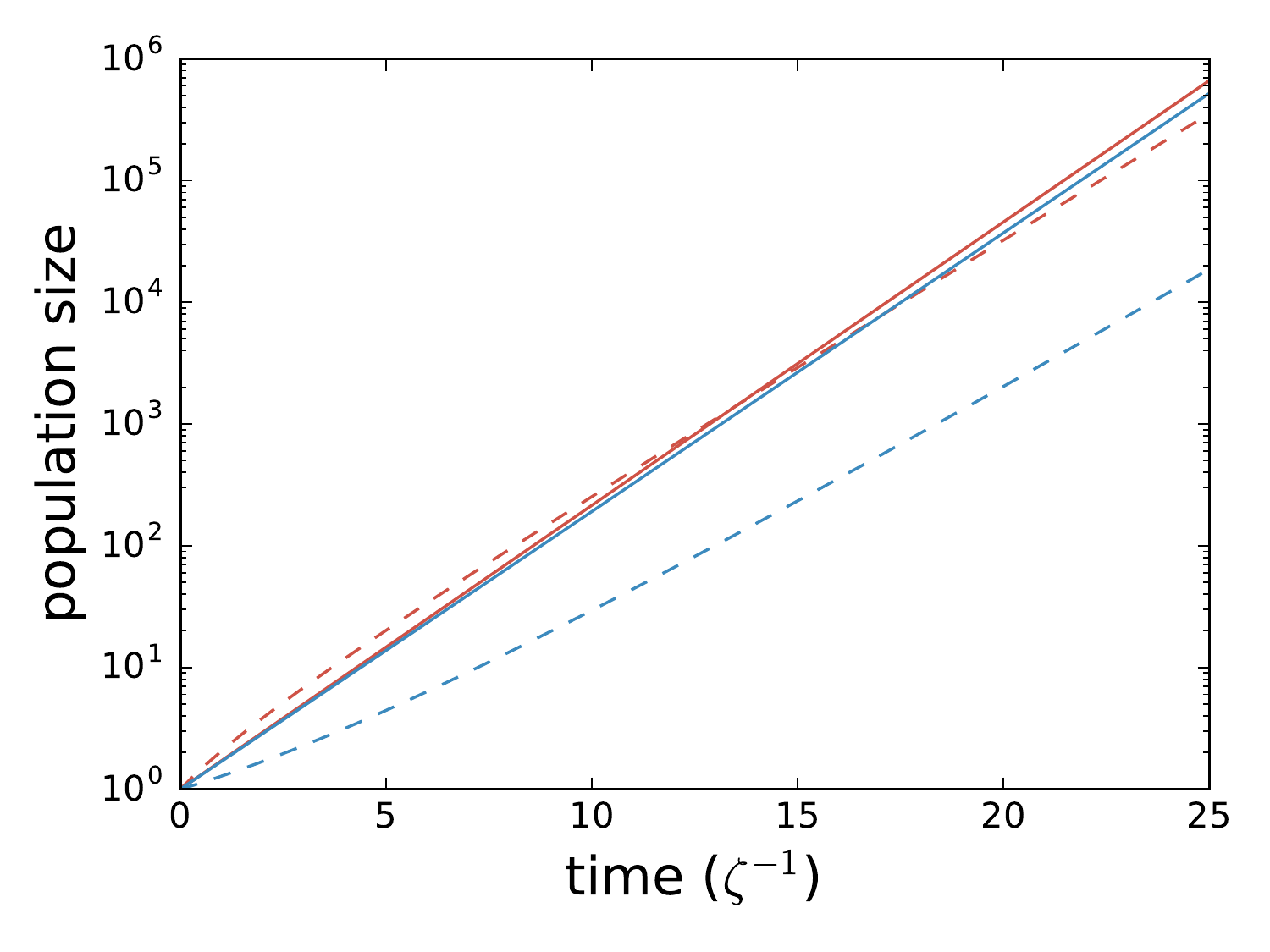} 
    \end{center}
    \caption{{\bf Growth trajectories under two scenarios.} Population size as a function of time for cell type 1 (red) and 2 (blue) under the scenario where each cell type
    maximizes its own growth rate (solid) and the scenario where both cell types maximize the growth rate of cell type 1 (dashed). The population of cells of type 1 in the
    competitive scenario overtakes that in the control scenario after it grows by a factor of $1420$.
\label{fig:gtraj} }
\end{figure}


\end{document}